\documentstyle[epsfig,aps,floats,eqsecnum,preprint]{revtex}

\def\laq{\raise 0.4ex\hbox{$<$}\kern -0.8em\lower 0.62ex\hbox{$\sim$}}
\def\gaq{\raise 0.4ex\hbox{$>$}\kern -0.7em\lower 0.62ex\hbox{$\sim$}}

\newcommand{\be}{\begin{equation}}
\newcommand{\ee}{\end{equation}}
\newcommand{\bees}{\begin{eqnarray}}
\newcommand{\ees}{\end{eqnarray}}
\newcommand{\pa}{\partial}
\newcommand{\co}{\nabla}
\newcommand{\tr}{{\rm Tr}}
\newcommand{\tg}{\tilde{g}}

\newcommand{\tR}{\tilde{R}}
\newcommand{\ophi}{\overline{\phi}}
\newcommand{\vophi}{\overline{\varphi}}

\begin{document}
\draft
\preprint{\vbox{\baselineskip=12pt
\rightline{CERN-TH/97-124}
\vskip0.2truecm
\rightline{hep-th/9706221}}}

\title{\Large\bf Classical Inhomogeneities in String Cosmology}
\author{A. Buonanno${}^{(a,b)}$, K.A. Meissner${}^{(a)}$
\footnote{Permanent address: Institute for Theoretical Physics,
Ho{\.z}a 69, 00-689 Warsaw, Poland.},
C. Ungarelli${}^{(a,b)}$ and G. Veneziano${}^{(a)}$}
\address{${}^{(a)}$ Theory Division, CERN, CH-1211 Geneva 23, Switzerland. \\
${}^{(b)}$ Istituto Nazionale di Fisica Nucleare, Sezione di Pisa,  
Pisa, Italy.}
\maketitle
\begin{abstract}
We generalize previous work on inhomogeneous pre-big bang cosmology
by including the effect of non-trivial moduli and
antisymmetric-tensor/axion fields.
The general quasi-homogeneous
asymptotic solution---as one approaches the big bang
singularity from perturbative initial data---is given and its
range of validity is discussed, allowing us to give a general
quantitative estimate of
the  amount of inflation obtained during the perturbative
pre-big bang era. The question of determining early-time
``attractors'' for generic pre-big bang cosmologies is also addressed,
and a motivated conjecture is advanced.
We also discuss S-duality-related features of the solutions, and
speculate on the way an asymptotic T-duality
symmetry may act on moduli space as one approaches the big bang.
\end{abstract}
\vskip 0.3truecm
\pacs{\tt PACS number(s): 11.25.-w, 98.80.Cq, 04.50.+h}

\section {Introduction}
\label{sec1}
After some pioneering work \cite{BV}, string cosmology
took a new turn with the realization
that, as a result of its duality symmetries \cite{duality,MV},
it naturally provides, even in the absence of potential
energy, standard (FRW) as well as inflationary
cosmologies. The crucial role
of a dynamical dilaton in providing inflation  then became clear.

This observation led to the idea of the so-called
pre-big bang (PBB) scenario \cite{MG1,MG2}
according to which the Universe started its evolution
from a very perturbative initial state,
i.e. from very weak coupling and very small curvatures.
It then inflated towards larger (space-time)
curvatures and coupling during the pre-big bang phase
and, possibly after a stringy epoch,
eventually made a transition (exit)
to the standard radiation-dominated era.

While early work concentrated mostly on homogeneous,
Bianchi I-type cosmologies,
and on small perturbations around them,  a
number of extensions of the original scenario
have been considered more recently, including some \cite{branes}
trying to incorporate the latest theoretical
developments in string theory. Within a more traditional
string theory framework, a more general
setting was recently considered \cite{inh}. In that
approach, given some initial data deeply inside the perturbative region,
but otherwise arbitrary,  their evolution is
followed towards the big bang singularity in the future.
It turns out that the evolution of fairly homogeneous initial
patches can be described analytically
and that a large fraction of those patches inflates,
becoming increasingly flat,
homogeneous and isotropic.
In the special case of exactly homogeneous and isotropic---but non
spatially flat---cosmologies,
 explicit solutions can be found \cite{Cop} \cite{TW}.
Also, some scepticism on the naturalness of the PBB picture arose~\cite{TW}.

If the above results are coupled to the assumption that a
``graceful exit'' does indeed take place
(see \cite{ex} for recent progress on this issue),
the usually {\it assumed} Big Bang
 conditions (a hot, dense and highly-curved state)
will be the outcome---rather than the starting
point---of inflation. Several interesting phenomenological
consequences of the whole scenario have been
worked out, particularly on the possibility of generating
an interesting spectrum of gravitational
waves \cite{tensor} and of cosmic magnetic fields \cite{em}.
Of particular relevance could be the recent observation
\cite{Copeland} that pre-big bang cosmology
can lead to a scale-invariant spectrum of axionic perturbations.

The purpose of this work is to improve on the results of \cite{inh} in 
various respects, as
explained in the following outline: in Sec. \ref{sec2},
we formulate the problem directly
in the string frame while adding extra
dimensions as well as the antisymmetric-tensor field
$B_{\mu \nu}$. We thus recover
the results of \cite{inh} and are able to extend them
to the case of quasi-homogeneous $B_{\mu \nu}$,
$g_{\mu \nu}$ and  $\phi$ fields. In Sec. \ref{sec3} we re-express
the four-dimensional case with torsion
and a single internal-space modulus in terms of the axion
field and construct the general
asymptotic solution for a quasi-homogeneous axion background.
We also look at the solutions
in the Einstein frame in order to expose,
as simply as possible, their
S-duality properties. In Sec. \ref{sec4} we discuss the limits of
validity of our asymptotic
solutions, from the point of view of both the breakdown of
the tree-level low-energy effective action and
of that of the gradient expansion. We are thus able to estimate
the duration of the PBB era and
the number of e-folds it generates, and to conjecture that the
far-past ``attractor'' of generic (negative-curvature)
PBB cosmologies coincides with the Milne Universe appearing in the explicit
solution of \cite{Cop}, \cite{TW}. Section \ref{sec5}
 contains some concluding remarks,
while we discuss  in the Appendix the structure of the
``momentum'' constraints, as well as their
solutions, in the particular case of $2+1$ dimensions.

\section {Pre-big bang cosmologies with quasi-homogeneous torsion}
\label{sec2}
\subsection{General lowest-order pre-big bang equations}
\label{sec2.1}
In this section we write down the classical equations of motion
of string
gravity at lowest order both in the $\alpha'$ and in the string-loop
expansion. Such an approximation should be valid at early
enough times since, in the pre-big bang scenario, the initial state of
the Universe is taken to be deeply inside the perturbative region.
Unlike in \cite{inh}, we work directly in the string frame,
where the physical interpretation of the solutions is
most immediate, and we consider the effect of internal dimensions
as well as that of a non-trivial antisymmetric-tensor field.

The low-energy effective string action in $D=d+1$ space-time
dimensions  is~\footnote{
We will use the signature $(-,+,+,\dots, +)$ and the following conventions
${\cal R}^\mu_{\,\,\,\,\nu \rho \sigma} = \Gamma^\mu_{\nu\sigma,\rho}
- \dots\,,\,\,{\cal R}_{\mu \nu} = {\cal R}^\rho_{\,\,\,\,\mu \rho \nu}$.
We indicate with ${\cal D}_\mu$ the covariant derivative
compatible with the  metric $g_{\mu \nu}$, while
${\nabla}_i$,  $R$ stand for the covariant derivative
and curvature obtained from the
(spatial) metric $g_{i j}$.}
\be
\label{ac}
{\cal S}_{\rm eff}^S = \frac{1}{2 \lambda_s^{D-2}}\,
\int d^D x\,\sqrt{-g} \, e^{-\phi}\,\left [ {\cal R} + g^{\mu \nu}
\pa_{\mu} \phi\,\pa_{\nu} \phi - \frac{1}{12}\,H_{\mu \nu \rho}\,
H^{\mu \nu \rho} \right ] \,,
\ee
where
\be
H_{\mu \nu \rho} = \pa_{\mu} B_{\nu \rho} +  \pa_{\nu} B_{\rho \mu}
+\pa_{\rho} B_{\mu \nu}\nonumber \,.
\ee
The equations of motion derived from the action (\ref{ac}) are well known: 
\bees
\label{a1}
&& {\cal R} + 2 g^{\mu \nu}\,{\cal D}_\mu\,{\cal D}_\nu \phi
- g^{\mu \nu}\,\pa_{\mu} \phi\, \pa_{\nu} \phi - \frac{1}{12}\,H_{\mu  
\nu \rho}
\,H^{\mu \nu \rho} = 0\,,\\
\label{a2}
&& {\cal R}_{\mu \nu} + {\cal D}_\mu\,{\cal D}_\nu \phi
- \frac{1}{4}\,H_{\mu \alpha \beta}\,H_\nu^{\,\,\,\,\alpha \beta}=0\,,\\
\label{a3}
&& \pa_{\nu} (\sqrt{-g}\,e^{-\phi}\,H^{\nu \rho \mu}) =0\,.
\ees
Invariance under general coordinate transformations and $B_{\mu \nu}
\rightarrow B_{\mu \nu} + \pa_{\mu} \Lambda_{\nu} - \pa_{\nu}
\Lambda_{\mu}$
allows us to bring the components $g_{0\mu}$ and $B_{0\mu}$
to the form
\be
g_{00}=-1\,,\quad \quad g_{0 i}= 0\,,\quad \quad B_{0i} = 0\,.
\ee
In this (synchronous) gauge we rewrite the above equations,
explicitly distinguishing time and space
derivatives.
To this end we introduce
\be
\chi^i_j = g^{ik}\,\pa_0 g_{kj},\quad\quad \ophi= \phi - \log(\sqrt{-g})\,,
\ee
and then  rewrite Eqs.~(\ref{a1})--(\ref{a3}) in the form
\be
\label{f1}
\ddot{\ophi} = \frac{1}{4}\,\tr(\chi^2) +
\frac{1}{4}\,g^{ik}\,g^{jl}\,\dot{B}_{ij}\,\dot{B}_{kl}\,,
\ee
\be
R^i_{\,j} + \frac{1}{2}\,\dot{\chi}^i_{\,j}
+ \nabla^i \nabla_j \phi - \frac{1}{2}\,\chi^i_{\,j}\,\dot{\ophi} -
\frac{1}{4}\,H^{ikl}\,H_{jkl} + \frac{1}{2}\,g^{im} g^{kl}\,\dot{B}_{mk}\, 
\dot{B}_{jl}=0 \,,
\label{f3}
\ee
\be
R + \frac{1}{4}\,\tr(\chi^2) +
\dot{\ophi}^2 - 2\ddot{\ophi} +
2\nabla^i\,\nabla_i \phi -
g^{i j}\,\pa_i \phi\,\pa_j \phi -
\frac{1}{12}\,H_{ikl }\,H^{ikl} +
\frac{1}{4}\,g^{ik}\,g^{jl}\,\dot{B}_{ij}\, \dot{B}_{kl}=0 \,,
\label{f4}
\ee
\be
\label{f5}
\pa_l \left ( e^{-\ophi}\,g^{l k}\,g^{i j}\, \dot{B}_{j k} \right )=0\,,
\ee
\be
\label{f6}
\pa_0 \left (e^{-\ophi}\,g^{i k}\,g^{j l}\, \dot{B}_{k l} \right )\,
-\pa_l \left (e^{-\ophi}\,H^{l i j}\right )=0\, ,
\ee
\be
\label{f2}
\frac{1}{2}\,\pa_k \chi_i^{\,\,k} - \frac{1}{4}\,(\pa_i g_{k
l})\,\chi^{k l} +
\pa_i \dot{\ophi}- \frac{1}{2}\,\chi_i^{\,\,j}\,\pa_j \ophi -
\frac{1}{4}\,\dot{B}_{j k}\,H_i^{\,\,j k}=0\,.
\ee
Equation~(\ref{f2}) represents the so-called ``momentum''
 constraints which, as such,
do not contain second-order time derivatives.
The remaining (so-called ``Hamiltonian'')
constraint is easily obtained by combining Eqs. (\ref{f1}) and (\ref{f4})
 and reads
\be
\label{f7}
R -\frac{1}{4}\,\tr(\chi^2) + \dot{\ophi}^2 +
2\,g^{i j}\,\nabla_i\,\nabla_j \phi -
g^{ij}\,\pa_i \phi\,\pa_j \phi -
\frac{1}{12}\,H_{ikl}\,H^{ikl} -
\frac{1}{4}\,g^{ik}\,g^{jl}\,\dot{B}_{ij}\,
\dot{B}_{kl}=0 \,.
\ee
Both (\ref{f2}) and (\ref{f7}) need only be imposed at a given time:
the  evolution equations then ensure
their validity at all times.

Equation (\ref{f1}) is independent of spatial gradients and
gives the important general result
$$
\ddot{\ophi}\ge 0 \,.
$$
Following \cite{inh}, our approach will consist in first solving
Eqs.~(\ref{f1})--(\ref{f6}) neglecting
spatial derivatives. As a result, the integration
``constants'' in the time-dependent solutions are actually functions
of the spatial
coordinates.  The ``momentum'' constraints (\ref{f2}) imply
$d$ relations among those
arbitrary functions, reducing their actual number
to the physically correct value.
The ``momentum'' constraints are
notoriously difficult to solve: in this paper, we will just assume
that they are somehow implemented.  In the Appendix we will discuss their
explicit form in the case $B_{\mu \nu}=0$ and $D=3$, where
solutions can be formally
given in terms of quadratures after a convenient choice of the spatial
coordinates has been made.

In the  following two sections we proceed to find quasi-homogeneous  
solutions,
by neglecting gradients in the equations. We
first discuss the case $B_{\mu \nu}= 0$, recovering,
in $D=4$, the results of \cite{inh}, and then consider
the general case.

\subsection{Quasi-homogeneous solutions with $B_{\mu \nu}= 0$}
\label{sec2.2}
Neglecting spatial gradients and comparing Eq.~(\ref{f1}) with
Eq.~(\ref{f7}) we obtain
\be
\label{sol3}
\ophi(t,\vec{x})=\ophi_0(\vec{x})-\log \left( 1-\frac{t}{t_0}
\right)\,,
\ee
{}while Eq.~(\ref{f3}) gives
\be
\dot{\chi}_{i}^{\,\,k}-\chi_{i}^{\,\, k}\,\dot{\ophi}=0 \,.  \label{r2}
\ee
Hence the solution for $g_{ij}$ reads
\be
\label{sol1}
g_{ij}(t,\vec{x})=\sum_a\,e^a_i(\vec{x})\,e^a_j(\vec{x})\,
\left( 1-\frac{t}{t_0}\right)^{2\,\alpha_a(\vec{x})}\,,
\quad \sum_a \alpha_a^2(\vec{x})=1\,,
\ee
where $e^a_i$ are arbitrary ``dreibein'' matrices and the constraint
on $\alpha_a$ implements Eq.~(\ref{f7}).
For the dilaton $\phi$ we get
\be
\label{sol2}
\phi(t,\vec{x})=\phi_0(\vec{x})-\gamma(\vec{x})\,\log \left( 1-\frac{t}{t_0}
\right)\,, \quad\quad \gamma(\vec{x})=1-\sum_a\alpha_a(\vec{x})  \,.
\ee
The solutions for $g_{i j}$ and $\phi$ are the most general ones; indeed
they depend on $d(d-1)$ arbitrary functions of space (after imposing the 
``momentum'' constraints and after gauge-fixing the spatial coordinates).
In $d=3$ these solutions reproduce those found in \cite{inh} by
transforming to the string frame the solutions found  in the Einstein  
frame.

Using $\pa_i \dot{\ophi}=0$ the ``momentum '' constraints, Eq.~(\ref{f2}), 
can be written in the form
\be
\pa_k(\chi^{\,\,k}_i\,e^{-\ophi})-\frac{1}{2}\,
e^{-\ophi}\,\tr[(G^{-1}\,\pa_i\,G)(G^{-1}\,
\dot{G})]=0\,.
 \label{r3}
\ee
Introducing the solutions~(\ref{sol3}) and (\ref{sol1}) in
Eq.~(\ref{r3}) we find
\be
\label{r4}
\sum_a\pa_k(e^{-\ophi}\,e^k_a\,e^a_i\,\alpha_a)
-e^{-\ophi}\,\sum_a\alpha_a\,(\pa_i e^a_k)\,e^k_a=0\,,
\ee
where we recall that it is sufficient to impose such constraints at any 
given time.

We are now able to discuss whether the asymptotic solutions
show some remnant of T-duality, which, in the case $B_{\mu \nu}=0$,  
reduces to
scale-factor-duality (SFD) and to its $Z_2^d$ generalization.
It is quite obvious from the form of the
solution that the transformation
\be
\label{SFD}
\alpha_a \rightarrow - \alpha_a, \quad \quad  e^a_i \rightarrow  
e^a_i, \quad \quad
\ophi_0(\vec{x}) \rightarrow \ophi_0(\vec{x})
\ee
generates, from any given solution, a dual one. It is less obvious to see
how a more general $Z_2^d$ transformation (i.e.
 $\alpha_a \rightarrow - \alpha_a$ for some $a$) can be implemented, since
the ``momentum'' constraint changes in a complicated way. Nevertheless, the case
of $d=2$ discussed in the Appendix  suggests that even the full
$Z_2^d$ duality group can be represented in the asymptotic solutions.
Understanding how that works in detail is beyond the scope of this paper. 
Such an understanding could shed new light on
some still outstanding problems \cite {Ricci} connected with the  
``non-Abelian''
generalization of T-duality \cite{NAD}.

\subsection{\bf  Quasi-homogeneous solutions in the presence of  
$B_{\mu \nu}$}
\label{sec2.3}
In the homogeneous case it is possible \cite{MV} to
recast the equations of motion in a form that is
manifestly covariant  under the global $O(d,d)$ group.
This certainly suggests that some trace of this symmetry should  also be
present  asymptotically in the inhomogeneous case.

We first write down the equations of motion in the form
\be
\label{q0}
\ddot{\ophi} = -\frac{1}{4}\,\tr(G^{-1}\,\dot{B}\,G^{-1}\,\dot{B}) +
\frac{1}{4}\,\tr(G^{-1}\,\dot{G}\,G^{-1}\,\dot{G}) \,,
\ee
\be
\label{q1}
\dot{\ophi}^2 + \frac{1}{4}\,\tr(G^{-1}\,\dot{B}\,G^{-1}\,\dot{B}) -
\frac{1}{4}\,\tr(G^{-1}\,\dot{G}\,G^{-1}\,\dot{G}) =
-2 \co^2 \phi +
(\co \phi)^2 - R + \frac{1}{12}\,H_{i kl}\,H^{i kl}\,,
\ee
\be
\label{q3}
\left [ \dot{B}\,G^{-1}\,\dot{B} + \dot{\ophi}\,\dot{G} -
G\,\pa_0(G^{-1}\,\dot{G}) \right ]_{i j} =
2\,R_{i j} + 2\,\co_i \,\co_j \phi - \frac{1}{2}\,
H_{i kl}\,H_{j}^{\,\,kl}\,,
\ee
\be
\label{q4}
\left [ G^{-1}\,\dot{\ophi}\,\dot{B} - G^{-1}\,\ddot{B} + G^{-1}\,
\dot{G}\,G^{-1}\,\dot{B}
+ G^{-1}\,\dot{B}\,G^{-1}\,\dot{G} \right ]^{i}_{\,\,j} = - e^{\ophi}\,
\pa_l ( e^{-\ophi}\,H^{l i k})\,g_{k j}\,,
\ee
where $G \equiv g_{i j}$ and $B \equiv B_{ij}$ are matrix
representations of the $d \times d$
spatial part of the metric and of the antisymmetric tensor.
We then introduce the usual $2d\times 2d$ matrices
\be
M= \left ( \begin{array}{cc}
G^{-1}, & - G^{-1}\,B \\
B\,G^{-1}, & G - B\,G^{-1}B\\
\end{array}
\right )\,,\quad\quad
\eta= \left ( \begin{array}{cc}
0 & \openone\\
\openone & 0\\
\end{array}
\right )\,,
\ee
define
\bees
U &\equiv& g_{i n}\,g_{j k}\,\pa_l(e^{-\ophi}\,H^{l n k }) \,,\\
V&\equiv& -e^{-\ophi}\,\left (2\,R_{i j}
+ 2\,\co_i \,\co_j \phi - \frac{1}{2}\,
H_{i kl}\,H_{j}^{\,\,kl}\right )\,,
\ees
and a new  $2d \times 2d$ matrix
\begin{equation}
\tilde{M}= \left ( \begin{array}{cc}
-G^{-1}V\,G^{-1}, & G^{-1}VG^{-1}B - G^{-1}U \\
U G^{-1} - BG^{-1}VG^{-1}, & V - UG^{-1}B+BG^{-1}VG^{-1}B-BG^{-1}U\\
\end{array}
\right )\,.
\end{equation}
The equations of motion, (\ref{q0})--(\ref{q4}), then become
\bees
&& \pa_t \left [e^{-\ophi}\,(M\,\eta\,\dot{M})\right ] =
M\,\eta\,\tilde{M}\,, \\
&& \ddot{\ophi} +\frac{1}{8}\,\tr (\dot{M}\,\eta\,\dot{M}\eta)=0\,, \\
&& \dot{\ophi}^2 + \frac{1}{8}\,\tr (\dot{M}\,\eta\,\dot{M}\eta) =
-R - 2\,\co_i \,\co^i \phi +
g^{i j}\,\pa_i \phi\,\pa_j \phi+
\frac{1}{12}\,H_{i kl}\,H^{i kl}\,.
\ees
If we neglect gradients we obtain
\bees
&& e^{-\ophi}\,(M\,\eta\,\dot{M}) = C(\vec{x})\,, \\
\label{oddeq}
&& C^T(\vec{x}) = -C(\vec{x}), \,\quad \quad
M\,\eta\,C(\vec{x}) = - C(\vec{x})\,\eta\,M\,,\\
&&  \ddot{\ophi} = \dot{\ophi}^2\,.
\ees
Defining the ``dilaton time'' $\tau$
\be
\tau = \int^t_0 e^{\ophi}\,d t^\prime =
-t_0\,e^{\ophi_0}\,\log\left (1 - \frac{t}{t_0} \right )\,,
\ee
the general solution of the equations of motion is
\bees
\label{mphisol}
M(t,\vec{x}) & = &\exp\left [-C(\vec{x})\,\eta\,\tau\right
]\,M_0(\vec{x})\,, \\
\ophi(t,\vec{x})& = & \ophi_0(\vec{x}) - \log \left (1-\frac{t}{t_0}  
\right )\,,
\ees
with
\be
\label{TrC}
\tr (C\,\eta)^2 =8\,e^{-2\ophi_0}/t_0^2\; .
\ee
These solutions represent an obvious generalization of the homogeneous
Bianchi I solutions given in \cite{MV}.
Disregarding gradients, equations of motion and their solutions
given above are manifestly covariant
under $O(d,d)$ transformations. On the contrary,  the
``momentum'' constraints, which become trivial in the absence of gradients,
cannot be expressed just in terms of the matrix $M$ and thus
appear to ``break'' T-duality.  More explicitly, they read
\be
\pa_k(\chi^{\,\,k}_i\,e^{-\ophi}) + 2 e^{-\ophi} \pa_i \dot{\ophi}
-\frac{1}{2}\,
e^{-\ophi}\,\tr[(G^{-1}\,\pa_i\,G)(G^{-1}\,\dot{G})]
-\frac{1}{2}\,\dot{B}_{j k}\,H_i^{\,\,\,jk}\,e^{-\ophi}=0\,,
\ee
i.e. in terms of  $M$
\be
\pa_k(\chi^{\,\,k}_i\,e^{-\ophi}) + 2 e^{-\ophi} \pa_i \dot{\ophi} +  
\frac{1}{4}\,
e^{-\ophi}\,\tr[\eta\,\dot{M}\,\eta\,\pa_i M]
- e^{-\ophi}\,g^{j l}\,g^{n k}\,(\pa_l B_{i n})\, \dot{B}_{k j}=0\,.
\ee
Using Eq.~(\ref{f5}) we get the final result
\be
\pa_k \left \{ e^{-\ophi} \left [(G^{-1}\,\dot{G})^{k}_{\,\,i} -
(G^{-1}\,\dot{B}\,G^{-1}\,B)^{k}_{\,\,i} \right ] \right \} +
\frac{1}{4}\,e^{-\ophi}\,\tr[\eta\,\dot{M}\,\eta\,\pa_i M] + 2
e^{-\ophi} \pa_i \dot{\ophi}
=0\,.
\label{momc}
\ee
In both Eq.~(\ref{f5}) and Eq.~(\ref{momc}) spatial derivatives
of upper entries of the matrix $e^{-\ophi}\,M\eta\dot{M}$ are present. 
They point to some remnants of $O(d,d)$ symmetry also in the
inhomogeneous case.

Hopefully, it is always  possible
to choose freely the matrix $C$, provided it satisfies
(\ref{oddeq}),(\ref{TrC}),
and then solve the momentum constraints with respect to  
$M_0(\vec{x})$. The
example of $D=3$
given in the Appendix supports this conjecture.
In this case the action of $O(d,d)$ on the (moduli)-space of
asymptotic solutions
is in principle well defined even though it is difficult, in practice,
to give it in an explicit form.

\section{Solutions with a quasi-homogeneous axion}
\label{sec3}
In this section we limit our attention to the case in which
all fields are independent of $n=D-4$ internal compact coordinates.
In this case
the components of the antisymmetric tensor $H^{\mu \nu \rho}$ with
$\mu, \nu, \rho = 0,1,2,3$ can be written in
terms of the pseudo-scalar axion $A$ as
\be
H^{\mu \nu \rho} \equiv E^{\mu \nu \rho \sigma} \,e^\varphi\,\pa_\sigma A\,,
\ee
where $E^{\mu \nu \rho \sigma}$ is the covariant, fully antisymmetric  
Levi-Civita tensor,
satisfying ${\cal D}_\alpha E^{\mu \nu \rho \sigma} = 0$.
We can then discuss, as an alternative to what we considered in the previous
section, the case of a quasi-homogeneous axion field. Because of the
duality relation between $A$ and $B_{\mu \nu}$, a quasi-homogeneous  
axion does
{\it not}
correspond to a quasi-homogeneous $B$-field. We shall carry out the
analysis first in the
string frame and then, in order to expose better S-duality-related  
features,
in the Einstein frame.

\subsection{String-frame description}
\label{sec3.1}
We  consider the possibility of a varying size (in ordinary space-time)
of the internal space by introducing a single modulus field $\beta$.
The reduced action following from Eq.~(\ref{ac}) becomes
\be
{\cal S}_{\rm eff}^S = \frac{1}{2 \lambda_s^2}\, \int d^4 x\,\sqrt{-g} \, 
e^{-\varphi}\,\left [ {\cal R} + g^{\mu \nu}
\pa_{\mu} \varphi\,\pa_{\nu} \varphi - \frac{1}{2}\,e^{2 \varphi}\,
g^{\mu \nu}\,\pa_\mu A\,
\pa_\nu A - n\,g^{\mu \nu}\,\pa_\mu \beta\,\pa_\nu \beta \right ] \,,
\label{string}
\ee
where $\varphi$  stands for the effective four-dimensional dilaton field 
$$
 \varphi = \phi  -n\,\beta\,.
$$
The equations of motion become
\bees
&& {\cal R} + 2 g^{\mu \nu}\,{\cal D}_\mu\,{\cal D}_\nu \varphi
- g^{\mu \nu}\,\pa_{\mu} \varphi\, \pa_{\nu} \varphi +
\frac{1}{2}\,e^{2 \varphi}\,g^{\mu \nu}\,\pa_\mu A
\,\pa_\nu A - n\,g^{\mu \nu}\,\pa_\mu \beta\,\pa_\nu \beta  = 0\,,\\
&& {\cal R}_{\mu \nu} + {\cal D}_\mu\,{\cal D}_\nu \varphi -
\frac{1}{2}\,e^{2 \varphi}\,\pa_\mu A\, \pa_\nu A
+ \frac{1}{2}\,g_{\mu \nu}\,e^{2 \varphi}\,g^{\rho \sigma}\,\pa_\rho A \,
\pa_\sigma A - n\,\pa_\mu \beta\,\pa_\nu \beta=0\,,
\ees
\bees
&& \pa_{\mu} (\sqrt{-g}\,e^{\varphi}\,g^{\mu \nu}\,\pa_\nu A) =0\,,\\
&& \pa_{\mu} (\sqrt{-g}\,e^{-\varphi}\,g^{\mu \nu}\,\pa_\nu \beta) =0\,.
\ees
Using $\chi_{ij} = \pa_0 g_{ij}$, we can rewrite these equations
in the synchronous gauge in the form
\be
\label{ll1}
-\ddot{\varphi} + \frac{1}{4}\,\tr(\chi^2) + \frac{1}{2}\,\dot{\chi} +
\frac{1}{2}\,g^{i j}\,e^{2 \varphi}\,\pa_i A\,\pa_j A +n\,\dot{\beta}^2=0\,,
\ee
\bees
\label{ll2}
&& R_{i j} + \frac{1}{2}\,\dot{\chi}_{i j} +
\frac{1}{4}\,(\chi \chi_{i j}-2 \chi_i^{\,\,k}\,\chi_{k j})
+ \nabla_i \nabla_j \varphi - \frac{1}{2}\,\chi_{i j}\,\dot{\varphi} -  
\nonumber \\
&& \frac{1}{2}\,e^{2 \varphi}\,\pa_i A\,\pa_j A - \frac{1}{2}\,e^{2  
\varphi}\,
g_{i j}\,\dot{A}^2 +
\frac{1}{2}\,g_{i j}\,g^{k l}\,\pa_k A \,\pa_l A
-n\,\pa_i\beta\,\pa_j \beta= 0\,,
\ees
\bees
\label{ll3}
&&  R + \frac{1}{4}\,\chi^2 +
\frac{1}{4}\,\tr(\chi^2) + \dot{\varphi}^2 - 2\ddot{\varphi}
-\chi\,\dot{\varphi} +\dot{\chi}+
2\,g^{ij}\,\nabla_i\,\nabla_j \varphi -
g^{ij}\,\pa_i \varphi\,\pa_j \varphi - \nonumber \\
&& \frac{1}{2}\,e^{2 \varphi}\,\dot{A}^2 + \frac{1}{2}\,e^{2  
\varphi}\,g^{i j}\,
\pa_i A \, \pa_j A +n\,\dot{\beta}^2 -n\,g^{i j}\,\pa_i \beta\,\pa_j  
\beta= 0\,,
\ees
\be
\frac{1}{2}( \nabla_j \chi_i^{\,\,j} - \pa_i \chi) +
\pa_i \dot{\varphi}- \frac{1}{2}\,\chi_i^{\,\,j}\,\pa_j \varphi -
\frac{1}{2}\,e^{2 \varphi}\,\dot{A}\,\pa_i A -n\,\dot{\beta}\,\pa_i  
\beta= 0\,,
\ee
\bees
&& \ddot{A} + \dot{\varphi}\,\dot{A} + \frac{1}{2}\,\chi\,\dot{A} =
g^{i j} \,\pa_i \varphi\,\pa_j A + g^{i j}\, \nabla_i \nabla_j A\,,\\
&& \ddot{\beta} - \dot{\varphi}\,\dot{\beta} +
\frac{1}{2}\,\chi\,\dot{\beta} =
-g^{i j} \,\pa_i \varphi\,\pa_j \beta + g^{i j}\, \nabla_i \nabla_j \beta\,.
\ees
Neglecting gradients, the  general solution of these equations
 is
\bees
&& g_{i j}(\tau,\vec{x}) = e^\varphi \, \sum_a e^{\,\,
a}_i(\vec{x})\,e^{\,\,a}_j(\vec{x})\,
\exp({2 \gamma_a(\vec{x})}\,\tau)\,, \\
&& e^{\varphi(\tau,\vec{x})} = \frac{C_1(\vec{x})}{K(\vec{x})}\,
\cosh [K(\vec{x})\,(\tau-\tau_0)]\,,\\
&& A(\tau,\vec{x}) = A_0(\vec{x}) \pm
\frac{K(\vec{x})}{C_1(\vec{x})}\,\tanh [K(\vec{x})\,(\tau-\tau_0)]\,, \\
&& \dot{\beta}(\tau,\vec{x}) =  
\frac{C_2(\vec{x})\,e^\varphi}{\sqrt{-g}}\,,\\
&& K(\vec{x}) = \sqrt{2\left [ (\sum_a \gamma_a(\vec{x}))^2
- \sum_a \gamma_a^2(\vec{x}) - n\,C_2^2(\vec{x})\right ]}\,,
\label{axstrsol}
\ees
where $\tau$ is the ``dilaton'' time
\be
\frac{d \tau}{d t} = e^{\vophi}\,, \quad \quad \vophi =
\varphi - \log(\sqrt{-g}) \,,
\label{dilt}
\ee
and $A_0(\vec{x}), C_1(\vec{x}), C_2(\vec{x}),\gamma_a(\vec{x})$ and  
$e_a^i(\vec{x})$ are arbitrary constants.

The above solutions, as well as the corresponding ones in the
Einstein frame presented in the
following subsection, generalize to the quasi-homogeneous case  the results
of Ref.~\cite{Cop}. For backgrounds with special symmetries similar solutions 
have been found in~\cite{KS}.    
Since the time dependence is implicit in the above
solutions (through Eq.~(\ref{dilt})), and their behaviour  is
similar to the one in the Einstein-frame,
we defer the discussion of both to Sec.~\ref{sec4}. We only note here  
that the dilaton field
has a singularity at both ends of the time evolution, even for a very small
axion field. However, before jumping to the conclusion
that we must face a strong-coupling regime in the far past, we
have to see what the actual range of validity of our approximations is.
This discussion too is postponed to Sec.~\ref{sec4}.

In order to expose the existence of an S-duality symmetry
connecting pairs of different solutions, we first consider the
same equations/solutions in the Einstein frame.

\subsection{Einstein-frame description}
\label{sec3.2}
The  Einstein-frame metric is obtained by
 the conformal transformation
\be
\label{tran}
\tilde{g}_{\mu \nu} = e^{-\varphi}\,g_{\mu \nu}\,.
\ee
The low-energy effective action with an axion and a modulus
field, Eq.~(\ref{string}), becomes
\be
\label{az}
{\cal S}_{\rm eff}^E = \int d^4 x\,\sqrt{-\tg} \,\left [
{\tR} -\frac{1}{2}\, \tg^{\mu \nu}\,
\pa_{\mu} \varphi\,\pa_{\nu} \varphi - \frac{1}{2}\,e^{2 \varphi}\,
\tg^{\mu \nu}\,\pa_{\mu} A\,\pa_{\nu} A -
n\,g^{\mu \nu}\,\pa_\mu \beta\,\pa_\nu \beta\right ] \,,
\ee
with equations of motion
\bees
&& \tR_{\mu \nu} - \frac{1}{2} \tg_{\mu \nu} \tR = \frac{1}{2}
\partial_{\mu} \varphi \partial_{\nu} \varphi -\frac{1}{4}\,\tg_{\mu\nu}
(\partial \varphi)^2  + \frac{1}{2}\,e^{2\varphi}\,
\partial_{\mu} A\, \partial_{\nu} A  
-\frac{1}{4}\,\tg_{\mu\nu}\,e^{2\varphi}\,
(\partial A)^2 +\nonumber \\
&& n\,\partial_{\mu} \beta \,\partial_{\nu} \beta
-n\,\frac{1}{2}\,\tg_{\mu\nu}(\partial \beta)^2\,, \\
&& \pa_\mu (\sqrt{-\tg}\,\tg^{\mu \nu}\,\pa_\nu \varphi) =
\sqrt{-\tg}\,e^{2 \varphi}\,
\tg^{\mu \nu}\,\pa_\mu A\,\pa_\nu A\,, \\
&& \pa_\mu ( \sqrt{-\tg}\,e^{2 \varphi}\,\tg^{\mu \nu}\,\pa_\nu A)=0 \,,\\
&& \pa_\mu ( \sqrt{-\tg}\,\tg^{\mu \nu}\,\pa_\nu \beta)=0 \,.
\ees
In the synchronous gauge $\tg_{0 0}=-1, \tg_{0 i}=0$ we obtain
\bees
\label{t6}
&& 2 \dot{\varphi}^2 + 4 n\,\dot{\beta}^2 +2 e^{2 \varphi}\,\dot{A}^2 +
\tr (\tilde{\chi}^2) - \tilde{\chi}^2 = 4\tR - \nonumber \\
&& 2 \tg^{i j}\,\pa_i \varphi\,\pa_j \varphi -
4 n\,\tg^{i j}\,\pa_i \beta\,\pa_j \beta - 2 \tg^{i j}\,
e^{2 \varphi}\,\pa_i A\,\pa_j A\,,
\ees
\be
\label{t3}
\ddot{\varphi} + \frac{1}{2}\,\tilde{\chi}\,\dot{\varphi} -
e^{2 \varphi}\,\dot{A}^2 =
\tg^{i j }\,\nabla_i \nabla_j \varphi - e^{2 \varphi}\,\tg^{i j}\,\pa_i  
A\,\pa_j A\,,
\ee
\be
\label{t1}
\dot{(\tilde{\chi}^i_{\,\,j})} +\frac{1}{2} \tilde{\chi}
\,\tilde{\chi}^i_{\,\,j} = -2\,
{\tR}^i_{\,\,j} +
\tg^{i k}\,\pa_k \varphi\,\pa_j \varphi+2n\,\tg^{i k}\,\pa_k  
\beta\,\pa_j \beta
 + e^{2 \varphi}\,\tg^{i k}\,\pa_k A\,\pa_j A\,,
\ee
\be
\label{t2}
\tilde{\chi}_{,i} - \tilde{\chi}_{i;\,j}^{\,\,j} =
- \dot{\varphi}\,\pa_i\varphi - 2n\,
\dot{\beta}\,\pa_i\beta
- e^{2 \varphi}\,\dot{A}\,\pa_i A\,,
\ee
\bees
\label{t4}
&& \ddot{A} + 2\dot{\varphi}\,\dot{A} + \frac{1}{2}\,\tilde{\chi}\,\dot{A} =
2\tg^{i j} \,\pa_i \varphi\,\pa_j A + \tg^{i j}\, \nabla_i \nabla_j A\,, \\
\label{t5}
&& \ddot{\beta} + \frac{1}{2}\,\tilde{\chi}\,\dot{\beta} =
\tilde{g}^{i j}\, \nabla_i \nabla_j \beta\,.
\ees
Disregarding  spatial gradients, Eq.~(\ref{t1}) can easily be solved,  
and we get
\be
\tilde{\chi} = \frac{2}{\tilde{t} - \tilde{t}_0}\,,
\ee
while the general solution can be written in the form
\bees
\label{z1}
&& \tg_{i j}(\tilde{t},\vec{x})=\sum_a\,\tilde{e}^a_i(\vec{x})\,
\tilde{e}^a_j(\vec{x})\,
\left(
  1-\frac{\tilde{t}}{\tilde{t}_0}\right)^{2\,\lambda_a(\vec{x})}\,,
\quad \sum_a \lambda_a(\vec{x})=1\,,\\
\label{z2}
&& e^{{\varphi}(\tilde{t},\vec{x})} = \frac{e^{\Phi_0(\vec{x})}}{2}\,
\left [ F(\vec{x})\,\left(1
-\frac{\tilde{t}}{\tilde{t}_0}\right)^{\tilde{q}(\vec{x})}+
\frac{1}{F(\vec{x})}\,\left(1
-\frac{\tilde{t}}{\tilde{t}_0}\right)^{-\tilde{q}(\vec{x})} \right
]\,,\label{p1} \\
&& {A}(\tilde{t},\vec{x}) = A_0(\vec{x})\pm 2 e^{-\Phi_0(\vec{x})}
\frac{ F(\vec{x})\left(
1-\frac{\tilde{t}}{\tilde{t}_0}\right)^{\tilde{q}(\vec{x})}}
{\left [F(\vec{x})\,\left(1
-\frac{\tilde{t}}{\tilde{t}_0}\right)^{\tilde{q}(\vec{x})}+
\frac{1}{F(\vec{x})}\,\left(
1-\frac{\tilde{t}}{\tilde{t}_0}\right)^{-\tilde{q}(\vec{x})}
\right ]}\,,\label{p2}\\
&& \beta(\tilde{t},\vec{x}) = \beta_0(\vec{x}) -
D(\vec{x})\,\log \left ( 1 - \frac{\tilde{t}}{\tilde{t}_0} \right )\,,
\ees
where $\Phi_0(\vec{x}), F(\vec{x}), D(\vec{x}), A_0(\vec{x})$ are
arbitrary functions of space and
\be
\tilde{q}(\vec{x}) = \sqrt{2}\,\sqrt{1 - \,\sum_a
\lambda_a^2(\vec{x}) - n\,D^2(\vec{x})}\,.
\ee
Some remarks on these solutions are in order. First of all, we still
have to impose the ``momentum constraints'' (\ref{t2}) (at any given
time) on these solutions.
The axion field, in the limit $\tilde{t} \rightarrow \tilde{t}_0$,
goes to the arbitrary
function $A_0(\vec{x})$. Since the dilaton field is increasing towards 
the singularity, terms like $e^{2\varphi}\,\nabla_i A\,\nabla^i A$
in the equations of motion may become important and can no longer be
disregarded. However, as we will see more accurately in
Sec. \ref{sec4}, we are not allowed to extrapolate the
solutions (\ref{z1})--(\ref{p2}) into the string phase, when
the coupling constant and/or the curvature (in string units) are of  
order $1$.
Imposing these limitations, it is possible to estimate the following  
behaviour
for the terms involving the axion field
\be
\frac{e^{2\varphi}\,\nabla_i A\,\nabla^i A}{\dot{\varphi}^2} \laq \,
{\cal O}\left (\frac{k^2_{\rm ph}}{M_s^2} \right )\,,
\ee
where $M_s$ is the string mass scale. Therefore, if we limit ourselves to 
energies much smaller than the string scale, it seems to be well
justified to neglect spatial
gradients in comparison with time derivatives.

Another important feature of the solution (\ref{z2})
is that it hits a strong coupling singularity in the far past,
as in the string frame.
As already mentioned, the discussion of this point is postponed to  
Sec. \ref{sec4}.

\subsection{S-duality}
\label{sec3.3}
Let us introduce the matrices $N \in SL(2,R)$:
\be
N= \left ( \begin{array}{cc}
e^\varphi, & e^\varphi\,A \\
e^\varphi\,A, & e^{-\varphi}+e^\varphi\,A^2\\
\end{array}
\right )\,,
\ee
and $J$:
\be
J= \left ( \begin{array}{cc}
0 & 1 \\
-1 & 0\\
\end{array}
\right )\,.
\ee
Neglecting the modulus field, the effective action (\ref{az}) becomes
\be
{\cal S}_{\rm eff}^E = \int d^4 x\,\sqrt{-\tg} \,\left [
{\tR} -\frac{1}{4}\,\tg^{\mu \nu}\,
\tr(J\,\pa_\mu N\,J\,\pa_\nu N) \right ] \,;
\ee
in particular, the ``momentum'' constraint, Eq.~(\ref{t2}), transform to
\be
\tilde{\chi}_{,i} - \tilde{\chi}_{i;\,j}^{\,\,j} =  
-\frac{1}{2}\,\tr(J\,\pa_0
N\,J\,\pa_i N)\,.
\ee
We note that the ``momentum'' constraint is manifestly
invariant under a generic transformation $\Theta$ of the group $SL(2,R)$
\begin{equation}
\Theta= \left ( \begin{array}{cc}
\delta & \gamma \\
\beta & \alpha\\
\end{array}
\right )\,,
\end{equation}
($\delta \alpha - \beta \gamma = 1$) acting on the backgrounds as follows
\be
\label{DT}
N \rightarrow \Theta^T N \Theta\,,
\ee
as  is easily checked using the property $\Theta^T\,J\,\Theta = J$.
The S-duality transformation (\ref{DT}) can be written
equivalently as
\be
S =  (A + i e^{-\varphi}) \rightarrow S' = \frac{\alpha S + \beta}
{\gamma\,S + \delta}\,.
\label{DT'}
\ee
In the same way as in~\cite{Copeland} and \cite{Fein},
we can relate, with an S-duality transformation of $e^\varphi$ and $A$,
the dilaton--vacuum solutions with a constant axion to particular
axion--dilaton solutions with a time-dependent axion.
Indeed, applying an S-duality transformation
to the dilaton--vacuum solutions with zero axion
\bees
\label{h1}
&& \tg_{i j}(\tilde{t},\vec{x})=\sum_a\,
\tilde{e}^a_i(\vec{x})\,\tilde{e}^a_j(\vec{x})\,
\left( 1-\frac{\tilde{t}}{\tilde{t}_0}\right)^{2\,
\lambda_a(\vec{x})}\,, \quad \sum_a \lambda_a(\vec{x})=1\,,\\
&& e^{\varphi(\tilde{t},\vec{x})}= e^{\varphi_0(\vec{x})}\,\left(
1-\frac{\tilde{t}}{\tilde{t}_0}\right)^{-\tilde{q}(\vec{x})}\,, \quad
\quad A(\tilde{t},\vec{x})=0\,,
\ees
we get the particular axion-dilaton solutions, with time dependent axion
\bees
\label{h2}
&& e^{\varphi^\prime(\tilde{t},\vec{x})} =  
\gamma^2\,e^{-\varphi_0(\vec{x})}\,
\left( 1-\frac{\tilde{t}}{\tilde{t}_0}\right)^{\tilde{q}(\vec{x})}+
\delta^2\,e^{\varphi_0(\vec{x})}\,\left(
1-\frac{\tilde{t}}{\tilde{t}_0}\right)^{-\tilde{q}(\vec{x})}\,,\\
\label{h3}
&& A^\prime(\tilde{t},\vec{x}) = \frac{\beta}{\delta} +
\frac{\gamma}{\delta}\,
\frac{ e^{-\varphi_0(\vec{x})}\,\left(
1-\frac{\tilde{t}}{\tilde{t}_0}\right)^{\tilde{q}(\vec{x})}}
{\left [\gamma^2\,e^{-\varphi_0(\vec{x})}\,
\left( 1-\frac{\tilde{t}}{\tilde{t}_0}\right)^{\tilde{q}(\vec{x})}+
\delta^2\,e^{\varphi_0(\vec{x})}\,\left(
1-\frac{\tilde{t}}{\tilde{t}_0}\right)^{-\tilde{q}(\vec{x})}\right ]}\,.
\ees
Unlike in the general case, in these particular
S-duality generated solutions the axion field approaches a
constant value as one goes
towards the singularity and  we do not face the problem
of a possible failure of the small-gradient expansion.
Also note that we can obtain Eqs.~(\ref{h2}),(\ref{h3}) directly from  
the general solutions
Eqs.~(\ref{p1}),(\ref{p2}) with the choice
\be
e^{\Phi_0(\vec{x})} = 2 \gamma\,\delta\,, \quad  F(\vec{x}) =
\frac{\gamma}{\delta}\,
e^{-\varphi_0(\vec{x})}\,, \quad A_0(\vec{x}) = \frac{\beta}{\delta}\,.
\ee

\section{Limits of validity of the approach}
\label{sec4}
The classical equations of motion considered in this paper and their
solution in the quasi-homogeneous case are valid only in a rather
restricted domain for the fields and their derivatives. Generally, there are
two restrictions concerning the use of the effective action coming
from string theory. The first one is that the coupling constant
should not be too large since otherwise the higher loop corrections
and non-perturbative effects (such as the dilaton
potential) start to be important.  The second limitation concerns the  
energy density (or if we prefer the curvature) which
should be smaller than the string scale. At higher energies/curvatures
massive modes of the string are excited and the whole picture based on the
massless background fields breaks down.

To these two very general restrictions we have to add, in our context, 
that of neglecting spatial derivatives. We have to check, a
posteriori, the range of
validity of this third approximation. The situation can be described
qualitatively as follows.

As one moves forward in time from fairly homogeneous initial conditions 
towards the singularity, the  approximation  of neglecting
spatial gradients becomes better and better. We should thus trust our  
asymptotic solutions
near the singularity as long as the other two general limitations
(on coupling and curvature) are
fulfilled. Unless we want to make assumptions about what happens
at strong coupling and/or curvature, these considerations limit the
duration of the PBB era
through an upper bound on its {\it end}.
However, in order to make a reliable
estimate of the total duration and of the number of e-folds, we also  
have to estimate the
time, in the past, at which the small-gradient approximation breaks
down, i.e. we have
to find the relevant constraint on the {\it beginning} of PBB inflation.
These questions  will be studied in Sec. \ref{sec4.1}.

The other important issue concerns the very early-time behaviour of our
solutions, much before  PBB inflation started. Here we enter a
regime in which spatial and time derivatives are of comparable importance.
Singularities or other features appearing in our asymptotic  solutions
in these regions cannot be trusted a priori. New techniques have to
be used in order to
find out what the  early-time behaviour of our solutions actually is. 
A first attempt to answer this
question was made in Ref. \cite{inh}. In Sec. \ref{sec4.2} we will
present some more results
on this topic and even motivate a conjecture on the nature of a
generic early-time
``attractor'' (going backwards in time) in the case of PBB solutions
with negative spatial curvature.

\subsection {Limits on perturbative  PBB inflation}
\label{sec4.1}
In order to estimate the duration and  number of e-folds of
the PBB phase, we will compute the next-to-leading-order corrections
to our asymptotic
formulae in the string frame and thus estimate the time at which
the small-gradient approximation breaks down. This instant will be taken as 
the beginning of the PBB phase.
Defining the variable
\be
W(t, \vec{x}) = e^{-\vophi(t, \vec{x})}\,,
\ee
Eqs.~(\ref{ll1})--(\ref{ll3}) with $A=0=\beta$ can be rewritten as
\bees
&&\ddot{W}(t, \vec{x}) = W(t, \vec{x})\,\Gamma(t, \vec{x}) \,,\\
&& {(\chi_i^{\,j}\,W(t, \vec{x}))}^. = -2W(t, \vec{x})\,
\Pi_i^{\,j}(t, \vec{x})\,,\\
&& \left ( \frac{\dot{W}(t, \vec{x})}{W(t, \vec{x})} \right )^2 =  
\frac{1}{4}\,
\tr(\chi^2) + \Gamma(t, \vec{x})\,,
\ees
where
\bees
&& \Gamma(t, \vec{x}) = -R + (\nabla \varphi)^2 -2g^{i j}\,\nabla_i
\nabla_j \varphi \,, \\
&& {\Pi}_i^{\,j}(t, \vec{x}) = R_i^{\,j} + \nabla_i\,\nabla^j \varphi \,.
\ees
Since in sufficiently isotropic regions the three-curvature $R$ goes like 
$(1-t/t_0)^{-2{\rm Max}\,\alpha_a}$ towards the singularity,
we are allowed to consider, in those regions, an
expansion in $(1-t/t_0)$ for the solutions of the above equations. We  
obtain
\bees
\label{corr1}
&&
W(t,\vec{x}) =   W_0(\vec{x})\, \left [ \left (1-\frac{t}{t_0}\right )
+ \int^t_{t_0} dt^\prime \,
\int^{t^\prime}_{t_0} dt^{\prime \prime} \left (1-\frac{t^{\prime  
\prime}}{t_0}\right )\,
\Gamma(t^{\prime \prime}, \vec{x}) \right ]\,, \\
\label{corr2}
&& \chi_i^j(t,\vec{x}) = \frac{2 W_0(\vec{x})}{W(t,\vec{x})}\, \left [ 
-\frac{1}{t_0}\,\sum_a \alpha_a\,e^a_i\,e_a^j +
 \int^t_{t_0} dt^\prime \left (1-\frac{t^\prime}{t_0}\right )\,{\Pi}_i^{\,j}
(t^\prime, \vec{x}) \right ]\,,
\ees
where all the terms in the integrals must be evaluated
on the leading solutions (\ref{sol1}),(\ref{sol2}).
It is easy to conclude that the small-gradients approximation
breaks down at a time $t_b$ such that
\be
\label{con1}
\left |1 - \frac{t_b}{t_0} \right | \sim \left
(\frac{L}{t_0} \right )^{1/(1-{\rm Max}\,\alpha_a)}\,,
\ee
where $L$ is the typical wavelength associated with the
inhomogeneities of the metric.
We observe that Eq.~(\ref{con1}) can also be written in the form
\be
\frac{R(t_b)}{\chi^2(t_b)} \sim {\cal O}(1)\,,
\ee
and then  choosing the initial time $t_i > t_b$ such that
\be
\label{con}
\frac{R(t_i)}{\chi^2(t_i)} \laq \,{\cal O}(1)\,,
\ee
we obtain a very natural condition to impose at the beginning of
the PBB inflationary phase.

We are now able to estimate the number of e-folds available in
the PBB cosmological model
with axion field, described in Sec. \ref{sec3.2}. The amount of inflation 
is commonly expressed in terms of the ratio of the comoving Hubble  
lengths at the
beginning and at the end of the inflationary phase
\be
Z = \frac{a(t_f)\,H(t_f)}{a(t_i)\,H(t_i)}\,.
\ee
Since we have  explicit solutions only in the Einstein frame,
we carry out the analysis
using Eqs.~(\ref{z1}) and (\ref{p1}), with $\beta=0$, and then
apply the transformation (\ref{tran})
in order to get $Z$ in the string frame. We obtain
\be
Z = \frac{\dot{\tilde{a}}(\tilde{t}_f) + \dot{\varphi}(\tilde{t}_f)/2}
{\dot{\tilde{a}}(\tilde{t}_i) + \dot{\varphi}(\tilde{t}_i)/2} \,,
\ee
where $\tilde{a}= (\sqrt{\tilde{g}})^{1/3}$.
We suppose that inflation starts at the time $\tilde{t}_i$ at which  the 
small-gradient expansion is still valid, Eq.~(\ref{con}),
and the dilaton field is near its
minimum, Eq.~(\ref{z2}), with $\dot{\varphi}_i >0$,
in order to have an expansion phase in the string frame.
We then get
\be
\left ( 1 - \frac{\tilde{t_i}}{\tilde{t_0}} \right ) \laq  
\,F^{-1/\tilde{q}}\,.
\ee
Let us address the question of setting a bound on the end of 
the perturbative PBB phase. Dilaton-driven inflation ends when either the string coupling
constant is of order $1$
or the (four-dimensional) curvature reaches the string scale  
$\lambda_s^{-2}$.
Imposing that $e^{\varphi_f} \laq \, 1$, we obtain (neglecting
numerical factors of  $O(1)$)
\be
\label{zn1}
Z \laq \,\exp \left ( - \frac{2}{3\tilde{q}}\,\Phi_0 \right )\,.
\ee
{}From ${\cal R}_f \laq \, \lambda_s^{-2}$ we get instead:
\be
\label{zn2}
Z \laq \, \left (\lambda_s^{2}  {\cal R}_i \right )^{-2/3(2-\tilde{q})}\,.
\ee
Note that in Eqs.~(\ref{zn1}) and (\ref{zn2}) the dependence on the
arbitrary constant $F$ drops out.

Combining the various results of this section and expressing
$\tilde{q}$ in terms of the
$\alpha_a$, we finally estimate the upper limit of the number of
e-folds during the PBB era as
\be
Z \laq \, \mbox{\rm Min} \left \{ \exp \left [-\Phi_0\,
\frac{(-1 + 1/3\,\sum_a \alpha_a)}{(-1+\sum_a \alpha_a)} \right ],
 \left (\lambda_s^{2}  {\cal R}_i \right )^{(\sum_a \alpha_a -3)/6}  \right \}\,.
\ee
This result generalizes to the inhomogeneous case that of \cite{TW}
and shows that, in order to solve the standard-cosmology
problems through  dilaton-driven PBB inflation, the beginning of the  
PBB era must indeed lie at very tiny coupling and curvature
(in string units), a point already made in~\cite{inh}.

We may ask at this point whether such initial data represent
an unreasonable amount of fine-tuning or, in any case, a larger
amount than what is needed in usual inflation. Such questions
 are hard to formulate
in precise physical terms. Rather than answering  semi-philosophical
questions,
we wish to underline a crucial difference between
our framework and the conventional one.

In conventional inflation the pre-inflationary era is supposed to lie
in the high-curvature quantum-gravity regime and one is thus facing
the problem of whether and how such a phase can prepare an ``initial'' state
that is fit to inflate (see \cite{infla,Muk} for a recent discussion).
By contrast, our pre-inflationary Universe
is very classical and described by the tree-level low-energy
string effective action.
As a consequence classical
solutions must contain (at least) two free parameters (moduli)
corresponding to as many transformations, which alter the action
just by a multiplicative
constant. These are:
\begin{itemize}
\item a constant \underline{shift} of the dilaton,
\item a constant \underline{rescaling} of the space-time coordinates.
\end{itemize}
The two moduli can be given the meaning of the value of the string coupling
and of the curvature (in string units) at the onset of the inflationary epoch
(here the transition from quasi-Milne to quasi-pre-big bang behaviour).
But these are exactly the two parameters that have to be very small
in order to ensure a long PBB era!

Thus the fine-tuning of string cosmology alluded to in \cite{TW}   
just consists
in choosing
these two moduli in a convenient region. Such a region has finite/infinite
extension, depending on the measure one adopts, but its boudaries are
certainly only one-sided for both moduli.

Another point to be recalled is that the scenario of \cite{TW}
is exactly homogeneous and thus the same alleged fine-tuning has to happen
everywhere in space. In our inhomogeneous Universe,
like in the chaotic scenario \cite{Linde}, it is sufficient
that a convenient patch develops initial conditions in the right
range of parameters in order that it undergoes sufficient inflation.
Other regions may not be as ``lucky'' and  will not experience a long inflationary era.
Unfortunately, we may not live long
enough to check whether this was the case, since those regions will end up
being much beyond our present horizon.

\subsection{Early- and late-time ``attractors'' of PBB cosmologies }
\label{sec4.2}
As explained at the end of the previous section, if we look
at earlier and earlier times, the approximation of neglecting
spatial gradients, rather than the effective action itself,
appears to become inadequate. While it looks impossible
to obtain analytic solutions in this regime, one can go
a long way towards understanding qualitatively how PBB-type solutions
behave towards very early times. For simplicity we shall restrict most
of our analysis to the case of $D=4$ and vanishing  
antisymmetric-tensor/axion.

\subsubsection{\bf Early-time fixed points}

Generalizing the results of \cite{inh} we shall now claim that, at
least in $D=4$, the only non-singular
early-time fixed point is the trivial Minkowski vacuum with a
constant dilaton.
The argument takes its simplest form in the E-frame where we refer to  
the case discussed in Sec. \ref{sec3.2}. A linear combination of
Eqs.~(\ref{t6}) to (\ref{t1}) gives
\be
\label{dot}
\dot{\tilde{\chi}} = -\frac{1}{2} \left [\tr( \tilde{\chi}^2) +
2\dot\varphi^2 + 4 n \dot{\beta}^2 +
2e^{2\varphi}\,\dot{A}^2 \right ]\,.
\ee
At a regular fixed point the l.h.s. of (\ref{dot}) vanishes and thus, 
since there cannot be  cancellations,  each term on the r.h.s.
has to vanish as well.

At this point we  use Eq.~(\ref{t4}) to argue that, modulo surface terms,
$\pa_i A =0$,
and Eq.~(\ref{t5}) to
conclude that also $\pa_i \beta=0$. Equation~(\ref{t1}) will finally give
${\tR}_i^{\,\,j}=0$, which, in three dimensions,
implies  flat space-time.

We conclude that initial data that do not come
from a singularity in the past must originate from the trivial vacuum of  
string theory.
The next question is whether the set of such initial data
has finite measure.

\subsubsection{\bf Heuristic criteria for a trivial early-time attractor}

Consider generic initial data subject to the constraint (satisfied,
in particular, for ${\tR} <0$)
\be
\Delta_E \equiv -2 {\tR} + \nabla_i\varphi\,\nabla^i \varphi +
2 n\,\nabla_i\beta\,\nabla^i \beta + e^{2\varphi}\,\nabla_i  
A\,\nabla^i A >0 \, ,
\ee
The obvious inequality (recall $\tr (\tilde{\chi}^2)/
\tilde{\chi}^2 \geq 1/3$):
\be
\dot{(\tilde{\chi}^{-1})} =
{1 \over 2 \tilde{\chi}^2}\,\left [\tr( \tilde{\chi}^2) +
2\dot\varphi^2 + 4 n \dot{\beta}^2 +
2 e^{2\varphi}\,\dot{A}^2 \right ] \ge {1 \over 6}\,,
\ee
holds at all times, while the inequality
\be
\dot{(\tilde{\chi}^{-1})} = {1 \over 2} -
{\Delta_E \over \tilde{\chi}^2} \le {1 \over 2}\,
\ee
holds at least initially.
The ratio ${\Delta_E/\tilde{\chi}^2}$ tends to decrease as we move
 forward in time
and to increase as we move backwards. It thus looks reasonable to
assume that, at least for a sufficiently isotropic situation,
${\Delta_E /\tilde{\chi}^2}$ does not change sign
 as we go towards the far past.
At the same time, the general
constraint ${\Delta_E / \tilde{\chi}^2} \le {1/3}$ is always
valid (see Eq. (\ref{t6})).

Under these assumptions we get:
\be
\label{cond}
\tilde{\chi}_i^{\,j} \sim \frac{c_i^{\,j}(\tilde{t},\vec{x}) }{\tilde{t}}\,,
\quad \quad 2\le c_i^{\,i} \le 6\,,\quad \quad 4 \le \tr(c^2) \le 36\,,\quad \quad
\tilde{t} \rightarrow - \infty \,.
\ee
We are now able to estimate the asymptotic behaviour
of the metric by first writing the formal solution
\be
\label{formal}
\tilde{g}_{ij}(\tilde{t}, \vec{x}) =
\left [T \exp\left( \int_0^{\tilde{t}} d \tilde{t}^\prime\,
\tilde{\chi}(\tilde{t}^\prime, \vec{x})\right)\right]_i^{\,k}
\tilde{g}_{kj}(0,\vec{x})
\ee
and by estimating the time-ordered integrals in (\ref{formal})
under the restrictions given in (\ref{cond}).
This gives:
\be
\label{asymptmetric}
\tilde{g}_{ij} \sim \tilde{e}_i^a(\vec{x})\,\tilde{e}_j^a(\vec{x})
\,(-\tilde{t})^{\bar{c}^{a}(\vec{x})}\,,
\ee
where $\bar{c}^a(\vec{x})$ is defined by:
\be
\bar{c}_i^{\,j}(\vec{x}) =  \tilde{e}_i^a(\vec{x})\,\tilde{e}^j_a(\vec{x})
\,\bar{c}^a (\vec{x})\,,\quad \quad 
\bar {c}_i^{\,j}(\vec{x}) = \lim_{\tilde{t} \rightarrow -\infty}\,\frac{1}{\tilde{t}}\,
\int_0^{\tilde{t}} d\tilde{t}^\prime\,\tilde{t}^\prime\, \tilde{\chi}_i^{\,j} 
(\tilde{t}^\prime, \vec{x})\;
\;.
\ee
Computing  the asymptotic behaviour
of ${\tR}$ from (\ref{asymptmetric})
we find a contradiction,
unless $\bar {c}^a =2$ for all $a$. In all other generic
 cases, i.e. barring special cancellations,
it is impossible to keep ${-{\tR}/\tilde{\chi}^2}$ (and a fortiori
 the ratio
${\Delta_E/\tilde{\chi}^2}$) bounded from above as $\tilde{t} \rightarrow -\infty$.
The only consistent solution is thus
\be
\label{milne}
\tilde{\chi}^j_{\,i}
= \frac{2 \delta^j_i}{\tilde{t}},\quad \quad  \dot{\varphi} =
\dot{\beta} =  \dot A = 0\,,
\ee
which is the Milne Universe \cite{Milne}. 
Using  previously given arguments, this leads
straight into the trivial vacuum as $\tilde{t} \to - \infty$.

For the homogeneous, constant-curvature cosmologies considered in
\cite{Cop},\cite{TW}, the fixed point just discussed is actually
reached  in the case of negative curvature
(i.e. of positive $\Delta_E$), while, for positive curvature, the
early-time ``attractor'' is singular.
We have checked that, instead, the Kantowski-Sachs cosmologies discussed
in Ref. \cite{KS}, being very particular and highly anisotropic, are able
to evade our general result.
 The argument given above leads us to
conjecture that the Milne Universe is
the Universal regular ``attractor'' for sufficiently isotropic generic 
initial data having
$\Delta_E >0$ everywhere. This conjecture will be further supported by  
performing a perturbative expansion of the general solution around
Milne space-time to show that it is a stable early-time fixed point. We
will also use the expansion near the singularity given in
Sec. \ref{sec4.1} in order to understand the flow of the solutions near  
the singularity.

\subsubsection{\bf Perturbative expansion around the early-time fixed point}

In this subsection we prove that the  Milne
metric with a  constant dilaton is indeed an
early-time ``attractor'', i.e. that it is stable against small
perturbations as we go backwards in time.

We recall that Milne's Universe \cite{Milne}
is actually isomorphic to a wedge of flat Minkowski space-time where
 constant-Milne-time hypersurfaces are constant-3-curvature
hyperboloids. In formulae the background
\bees
\label{atrmetr}
&& ds^2=-d\tilde{t}^2+\tilde{t}^2\,\left(\frac{dr^2}{1+r^2}+r^2d\Omega^2\right)\,,\\
&& \varphi={\rm const.}\,, \quad \quad \tilde{\chi}^i_{\,j}=\frac{2\delta^i_j}{\tilde{t}}\,,
\quad \quad {\tR}^i_{\,j}= -\frac{2\delta^i_j}{\tilde{t}^2}\,,
\ees
can be brought to Minkowski's form by the transformation:
\be
\label{MilMin}
-\tilde{t} = \sqrt{\tau^2-\rho^2}\,, \quad \quad  r = {\rho \over  
\sqrt{\tau^2-\rho^2}}\,, \quad \quad
\tau \le 0\,,\, \quad \quad  \rho^2 \le \tau^2\,,
\ee
where $\rho$ and $\tau$ are the Minkowskian coordinates.  
Given the basic triviality of Milne's Universe it is not surprising
that the exact field equation for $\varphi$ in Einstein's
frame, Eq.~(\ref{t3}), can be completely solved in such a background.
The result (see for instance \cite{TS}) reads:
\bees
\label{phiMilne}
&& \varphi = \varphi_0 + {T_0 \over \tilde{t}} \sum_{l,m} Y_{l m}(\Omega)\, 
C_{l m}(r,\tilde{t})\,, \nonumber \\
&& C_{l m}(r,\tilde{t}) = \int_{-\infty}^{\infty} dp~ b_{p l m}~
{\exp}(-ip\log(-\tilde{t}/{T_0}))~{\cal{P}}_{p l}(r) \,,\\
&& {\cal{P}}_{p l} = { P_{ip -1/2}^{-1/2 -l}(\sqrt{1+r^2})\over \sqrt{r}}\,,
\ees
where $Y_{l m}(\Omega)$ and ${P}_\mu^\nu$ are the usual spherical harmonics 
and associated Legendre functions respectively. The two classical moduli discussed
earlier show up in the general solutions as 
$\varphi_0$, an arbitrary constant, and $T_0$, an arbitrary overall
normalization parameter with the dimensions of
time (recall that $\varphi$ and $r$ are dimensionless). 
In particular the ``s-wave'' ($l=0$) contribution takes the 
form (with a slightly redefined coefficient $\bar{b}_{p00}$):
\be
\label{swave}
\varphi = \varphi_0 + {T_0 \over r(-\tilde{t})} 
\int_{-\infty}^{\infty} dp~ \frac{\bar{b}_{p 0 0}}{i p} ~
\left\{ e^{-ip [\log(-\tilde{t}\,(\sqrt{1+r^2} + r)/T_0)]} -
e^{-ip [\log(-\tilde{t}\,(\sqrt{1+r^2} - r)/T_0)]} \right \}\,.
\ee
The qualitative study of the general
solution is greatly helped by recalling
the large and small $r$ limit of the function ${\cal{P}}_{p l}$
appearing in (\ref{phiMilne}), i.e.
\be
{\cal{P}}_{p l}(r) \sim r^l \quad \quad r \rightarrow 0 \,, \quad \quad 
{\cal{P}}_{p l}(r) \sim \frac{1}{r}\,\Re{\it e}\left ( \frac{e^{ip\log 2r}\,\Gamma(ip)}
{\Gamma(1+l+ip)} \right )
 \quad \quad  r \rightarrow \infty\,.
\ee
Clearly, $\varphi \rightarrow \varphi_0$ as $\tilde{t}=-\infty$, and
the background goes, as expected, to the trivial vacuum.
Given a properly normalized distribution of the Fourier
coefficients $\bar{b}_{p 0 0}$, $(\dot{\varphi}/\chi)^2 \,\laq \, (T_0/\tilde{t})^2$
remains very small for $\tilde{t} \ll -T_0$. The same is true 
for the spatial derivatives of $\varphi$ and it is also expected to be the case 
for the fluctuations of the metric itself, since
they are similarly behaved~\cite{TS}. This confirms that Milne's Universe is 
an increasingly accurate solution to Eq.~(\ref{t6}) as $\tilde{t} \rightarrow -\infty$.

However, the back-reaction from the fluctuations of the dilaton and the metric 
becomes $O(1)$ at $\tilde{t} \sim -T_0$ and
$r\, \laq\, 1$. As a result, we expect the Milne background to turn quickly  
into a pre-big bang behaviour at least in regions where the spatial curvature 
is still negative. For $\tilde{t}\,\gaq -T_0$ the approximate solutions will be 
those given in Sec.~\ref{sec3} up to a redefinition of the time at which the singularity 
occurs. Typically, a whole region of proper size $O(T_0)$ (the Hubble horizon at $\tilde{t}=-T_0$)
is expected to undergo inflationary behaviour. We plan to investigate this
phenomenon numerically in a future publication.

Let us now study the perturbative expansion around Milne's Universe. 
The perturbed Eqs.~(\ref{t6})--(\ref{t1}) read
\bees
\label{perteq1}
(\dot{\delta\varphi})^2+\frac2{\tilde{t}}\,\delta \tilde{\chi}+
\dot{(\delta \tilde{\chi})}+
\frac{1}{2}\,\delta \tilde{\chi}_{\,i}^{j}\,\delta \tilde{\chi}_{\,  
j}^{i}&=&0\,,\\
\label{perteq3}
\dot{(\delta \tilde{\chi}^i_{\,j})}+\frac3{\tilde{t}}\,\delta
\tilde{\chi}^i_{\,j}+\frac1{\tilde{t}}\,\delta^i_j
\delta \tilde{\chi}
+2\delta{\tR}^i_{\,j}+ \tilde{g}^{i k}\, \pa_k \delta
\varphi\,\pa_j \delta \varphi&=&0\,,\\
\ddot{(\delta\varphi)}+ \frac{1}{2}\,\delta  
\tilde{\chi}\,\dot{(\delta\varphi)} +
\frac3{\tilde{t}}\,\dot{(\delta\varphi)}-\frac1{\tilde{t}^2}\,\delta  
(\Delta \varphi)&=&0\,,
\label{perteq2}
\ees
where $\Delta$ denotes the three-dimensional Laplacian built up from the
three-dimensional metric appearing in (\ref{atrmetr}).
We will see below that it is consistent to disregard the second term
in Eq.~(\ref{perteq2}) and the term that comes out from
the variation of the Laplacian. Therefore Eq.~(\ref{perteq2}) is  
decoupled from the rest and
can be easily solved.

\begin{figure}
\centerline{\epsfig{file=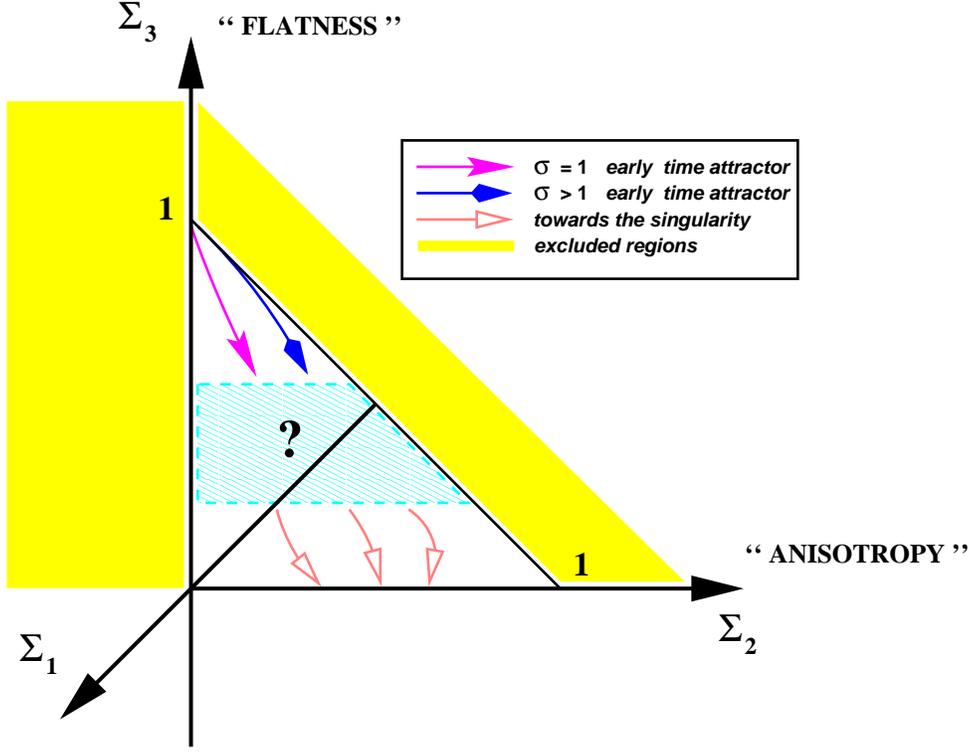,width=0.6\textwidth,angle=-90}}
\vspace{0.5cm}
\caption{\sl Approximate flow diagram from the early-time attractor to  
the singularity.}
\label{fig}
\end{figure}

For simplicity, we will restrict ourselves to the ``s-wave'' case and we set $T_0=1$. 
Using Eq.~(\ref{swave}) we get
\be
\label{gendp}
\delta\varphi=\frac{1}{r (-\tilde{t})}\,\left [{\cal G}
\left(-\tilde{t}(r+\sqrt{1+r^2})\right)-
{\cal G}\left(-\tilde{t}(-r+\sqrt{1+r^2})\right)\right]\,,
\ee
where ${\cal G}$ is a priori an arbitrary function.
However, for a sufficiently well behaved distributions 
of Fourier coefficients $\bar{b}_{p 0 0 }$, 
the function ${\cal G}$ is bounded by a constant at large $t$ 
(and fixed $r$) and we can write, up to oscillatory factors,
\be
\label{u1}
\delta\varphi=\frac{{\cal F}(r)}{(-\tilde{t})^\sigma}\,,
\quad \quad \sigma >1\,, \quad \quad \tilde{t} \rightarrow - \infty\,.
\ee
Eqs.~(\ref{perteq1}) and (\ref{perteq3}) are solved by the ansatz
\be
\delta\tilde{\chi}^i_j= \frac{{\cal A}^i_{\,j}}{\tilde{t}^2}+
\frac{{\cal B}^i_{\, j}}{(-\tilde{t})^{1+2\sigma}}
+\frac{{\cal C}^i_{\,j}}{\tilde{t}^3} + \ldots \,,
\label{u2} 
\ee
and using Eq.~(\ref{perteq2})  we get
\be
\tr\,{\cal B} = \frac{\sigma^2 {\cal F}^2}{2\sigma-1},\ \ \ \ \   
\tr\,{\cal C} =
\frac12\,\tr ({\cal A}^2)\,.
\label{trbtrc}
\ee
We now come to a qualitative analysis of the phase space of the trajectories 
in the range of validity of our solutions, either near the early-time  
``attractor'' or towards the singularity.
\vskip 0.2truecm
\paragraph{\bf  Behaviour near the early-time ``attractor''.} 
In the Einstein frame the ``Hamiltonian'' constraint reads
\be 
\Sigma_1 + \Sigma_2 + \Sigma_3 =1\,,
\ee
where
\be
\Sigma_1\equiv\frac{3\,\dot{\varphi}^2}{\tilde{\chi}^2}\,,\quad \quad 
\Sigma_2\equiv \frac{3\, \tr(\tilde{\chi}_T^2)}{2\,\tilde{\chi}^2}\,,\quad \quad 
\Sigma_3 \equiv 3\left[\frac12-\dot{(\tilde{\chi}^{-1})}\right] = 
\frac{3 \Delta_E}{\chi^2}\,,
\ee
and for any matrix we have defined
\be
({S}_T)^i_{\,j} = S^i_j-\frac13\delta^i_j\,\tr S\,.
\ee
This allows us to draw a ``flow diagram'' in the plane 
shown in Fig.~\ref{fig}, where 
to simplify the presentation we constrain $\sigma$ to be in the region
$1<\sigma<3/2$. The result for larger values of $\sigma$ is similar.
The individual contributions from the
perturbed solutions, Eqs.~(\ref{u1}) and (\ref{u2}), are (we neglect terms  
with power $1/\tilde{t}$ higher than $2$):
\bees
\Sigma_1&\simeq&
\frac{\sigma^2 {\cal F}^2}{12\,\tilde{t}^{2\sigma}}\,,\\
\Sigma_2 &\simeq&
\frac{\tr({\cal A}_T^2)}{24\,\tilde{t}^2}\,,\\
\Sigma_3&\simeq& 1-\frac{\sigma^2 {\cal F}^2}{12\,\tilde{t}^{2\sigma}}
-\frac{\tr({\cal A}_T^2)}{24\,\tilde{t}^2}\,.
\label{contr}
\ees
As can be seen from this result, there are two different behaviours of
these contributions when $\tilde{t}\to -\infty$:
for $\sigma=1$, $\Sigma_1, \Sigma_2$ and $(1-\Sigma_3)$ 
are of the same order, while for $1<\sigma<3/2$,
$\Sigma_2$ and $(1-\Sigma_3)$  are dominant.
In Fig.~\ref{fig} the first case, $\sigma=1$, is represented by 
trajectories that can have any angle (less than $\pi/2$)
with respect to the $\Sigma_3$ axis, in the second  case the trajectories 
are tangent to the straight line $\dot{\varphi}=0$. 

For $\sigma>3/2$ the behaviour of the contributions is the same as in
the $1<\sigma<3/2$ region, but there are higher-order terms to all
of them.
\vskip 0.2truecm
\paragraph{\bf Approaching the singularity.} 
In Sec. \ref{sec4.1} we have derived in the string frame the  
next-to-leading solutions
in the small-gradient expansion.
These solutions must satisfy the ``Hamiltonian''
constraint (\ref{f7}), which can be put in the form
\be
\Upsilon_1 + \Upsilon_2+\Upsilon_3=1\,,
\ee
where
\be
\Upsilon_1 \equiv \frac{\chi^2}{12\dot{\vophi}^2}\,,\quad \quad
\Upsilon_2 \equiv \frac{\tr({\chi}_T^2)}{4\dot{\vophi}^2}\,, \quad \quad 
\Upsilon_3 \equiv \frac{\Gamma}{\dot{\vophi}^2}\,.
\ee
To evaluate the slope of the trajectories in the ``flow diagram'', we  
will use Eqs.~(\ref{corr1}) and (\ref{corr2}). Hence we get
\bees
\label{var1}
\Upsilon_1 &\simeq&\frac{1}{3}\,(\sum_a \alpha_a)^2+ \frac{2}{3}\,{\cal I}\,,\\
\label{var2}
\Upsilon_2 &\simeq&1-\frac{1}{3}\,(\sum_a \alpha_a)^2- \frac{2}{3}\,{\cal  
I}-\Gamma\,(t_0-t)^2\,,\\
\label{var3}
\Upsilon_3 &\simeq& \Gamma\,(t_0-t)^2\,,
\ees
where
\bees
{\cal I}(t, \vec{x}) &=& \int^t_{t_0} dt^\prime \, (t_0-t^\prime)\, \left \{
\left [ (\sum_a \alpha_a)^2 - \sum_a \alpha_a \right ] \,
(-R ) +  \right. \nonumber \\
&& \left. \left [ (\sum_a \alpha_a)^2 + \sum_a \alpha_a \right ] \,  
(\nabla \varphi)^2- 
2  (\sum_a \alpha_a)^2 g^{ij}\,\nabla_i \,\nabla_j \varphi \right \}\,.
\ees
Using the relations
\bees
&& \tilde{\chi}_{\,j}^i = e^{\varphi/2}\,(\chi_{\,j}^i-\delta^i_{\,j}
\,\dot{\varphi})\,,\\
&& \tilde{R} = e^{\varphi}\,\left [ R + 2g^{i j}\,\nabla_i\,\nabla_j  
\varphi
- \frac{1}{2}\,g^{i j}\,\pa_i\varphi\,\pa_j \varphi \right ]\,,
\ees
we can express the Einstein-frame quantities $\Sigma_1$, $\Sigma_2$ and 
$\Sigma_3$ in terms of the string frame's $\Upsilon_1$, $\Upsilon_2$ and $\Upsilon_3$.
We get, in the super-inflationary PBB case,
\vskip 0.2truecm
\bees
\label{j1}
\Sigma_1 &=& \frac{3(1 + \sqrt{3 \Upsilon_1})^2}{(3 + \sqrt{3  
\Upsilon_1})^2}
\simeq \frac{3(1 - \sum_a \alpha_a)^2}{(3 -\sum_a \alpha_a)^2} -
\frac{12(1-\sum_a \alpha_a)}{(\sum_a \alpha_a)(3-\sum_a  
\alpha_a)^3}\,{\cal I}\,,\\
\label{j2}
\Sigma_2 &=& \frac{6 \Upsilon_2}{(3 + \sqrt{3 \Upsilon_1})^2}
\simeq \frac{2(3 - (\sum_a \alpha_a)^2)}{(3 -\sum_a \alpha_a)^2} +
\frac{12(1-\sum_a \alpha_a)}{(\sum_a \alpha_a)(3-\sum_a \alpha_a)^3}
\,{\cal I}-\frac{6 \Gamma\,(t_0-t)^2}{(3 -\sum_a \alpha_a)^2} \,,\\
\label{j3}
\Sigma_3 &=& \frac{6 \Upsilon_3}{(3 + \sqrt{3 \Upsilon_1})^2}
\simeq \frac{6 \Gamma\,(t_0-t)^2}{(3 -\sum_a \alpha_a)^2}\,.
\ees
\vskip 0.2truecm
Note that in a super-inflationary cosmological model ($\sum_a\alpha_a<0$) 
with negative three-curvature, if we are in a sufficiently isotropic  
region where
$\sum_a\alpha_a<-1$ we have globally ${\cal I} <0$ and then the  
behaviour in time is
\vskip 0.2truecm
\bees
\Sigma_1 &\simeq& \frac{3(1 - \sum_a \alpha_a)^2}{(3 -\sum_a \alpha_a)^2}-
\Xi^2\,\log^2\left (1-\frac{t}{t_0} \right )\,
\left (1-\frac{t}{t_0} \right )^{2-2{\rm Max}\,\alpha_a}\,,\\
\Sigma_2 &\simeq&  \frac{2(3 - (\sum_a \alpha_a)^2)}{(3 -\sum_a  
\alpha_a)^2}+
(\Xi^2-\Theta^2)\,\log^2\left (1-\frac{t}{t_0} \right )\,
\left (1-\frac{t}{t_0} \right )^{2-2{\rm Max}\,\alpha_a}\,,\\
\Sigma_3 &\simeq& \Theta^2\,\log^2\left (1-\frac{t}{t_0} \right )\,
\left (1-\frac{t}{t_0} \right )^{2-2{\rm Max}\,\alpha_a}\,.
\ees
\vskip 0.2truecm
Hence, we can conclude that $\delta\Sigma_1, \delta\Sigma_2$ and $\Sigma_3$ are
all of the same order in the limit $t \rightarrow t_0$, and the  
trajectories
of the solutions in the flow diagram (see Fig.~\ref{fig}) can have  
any slope
relative to the $\Sigma_2$ axis, depending on the values of $\Xi$ and  
$\Theta$.

\section {Summary and Discussion}

We have been able to extend previous work by showing that, even in  
the presence
of an antisymmetric-tensor/axion background, or of  
internal-dimension moduli,
 pre-big bang type
inflation emerges naturally in string theory from generic
 initial perturbative data. Reasonably smooth initial patches inflate and
keep becoming increasingly
homogeneous and (spatially) flat, at least as long as the low-energy 
tree-level effective action description is valid.

We were able to estimate the duration of the perturbative
pre-big bang phase and to show that it depends on two (arbitrary) classical
moduli. A sufficient amount of inflation requires these moduli
to be both bounded from above, something we do not believe has much
to do with
the concept of fine-tuning. The question remains open of whether higher-order 
($\alpha'$ or loop) corrections
might deform the classical moduli space and allow a
prediction for the duration of perturbative pre-big bang inflation.

We have also analysed the behaviour of our solutions towards the far past
and argued in favour of the existence of a rather large basin of
attraction towards a Milne-type Universe with trivial dilaton,
axion and moduli. Since Milne's Universe is equivalent to
(a wedge of) Minkowski space-time, such a state is nothing but
a disguised form of the exact perturbative
vacuum of string theory. This result (which should be established on more
rigorous grounds) leads to a striking confirmation of the viability
of the basic pre-big bang postulate, stating that
 the Universe started its evolution
from the trivial vacuum of string theory.

As we have shown in Section 4,
such a state, being an early-time attractor,  actually becomes
a repulsor as we move forward in time, i.e.
is classically unstable with respect to small fluctuations of the metric
and of the dilaton--axion system.
The generic cosmologies that spring out of the trivial vacuum consist of a
quasi-Milne era, followed by an inflationary quasi-homogeneous
pre-big bang era. The value of
the string coupling and of the spatial curvature at the transition between
the two phases are the two above-mentioned classical  moduli.

Quite possibly,  the most generic kind of string
cosmology will be quite inhomogeneous
in a global sense, since, a priori, the two moduli may take different values
in different regions of space. As in chaotic inflation
\cite{Linde}, homogeneity is a local property
valid up to some scale determined by the size of the original
patch, which gave rise to our observable Universe, and by the
amount of inflation it suffered.

We finally recall that, once pre-big bang behaviour sets in,  
primordial vacuum
fluctuations are parametrically amplified.  Equivalently, in a  
particle-physics
language, massless quanta are
copiously produced by the time-dependent backgrounds. By the time
the string coupling has grown to about its present value, these  
quanta are able
to dominate the energy and to lead the Universe straight into the hot  
big bang
era \cite{PBBB},\cite{BMUV2}.

Many points are still unclear throughout the picture, and much work
is still needed, both on the topics discussed in this paper
and on the issue of the transition from the pre-big bang
to the FRW phase (the exit problem). However,
the possibility that the
hot big bang conditions---which we know to have prevailed some
15 billion years ago---simply emerged from the basic instability of the trivial (i.e.
cold, empty, flat, free) vacuum of string theory appears to be gaining
further credibility from the results  reported here.

\section*{Acknowledgements}
One of us (G.V.) acknowledges interesting discussions
with J.D. Barrow on the generic chaotic behaviour of
solutions near the singularity in pure Einstein's gravity,
with S. Mukhanov on the question of fine-tuning in ordinary
and pre-big bang inflation, and with M. Turner and E.
Weinberg on their work.
We are also grateful to G. Pollifrone for useful conversations.
This work was supported in part by the EC contract No. ERBCHRX-CT94-0488.
A.B. and C.U. are partially supported by the University of Pisa.

\label{sec5}
\appendix
\section{}
\label{app}
We will restrict ourselves to the case without axion and modulus fields.

\subsection*{Momentum constraints for $d=2$: Einstein frame}
In the $2+1$ dimensional case we can choose the spatial
coordinates in such a way as to make the ``zweibeins'' diagonal.
Equations ~(\ref{z1}) and (\ref{z2}) then become
\bees
&& \tg(\tilde{t},x,y) = e^{\tilde{\beta}_i(x,y)}\,\delta_{i j}\,
\left(1-\frac{\tilde{t}}{\tilde{t}_0}\right)^{2\,\lambda_i(x,y)}\,,\quad
\lambda_1(x,y)+\lambda_2(x,y)=1\,, \\
&& \varphi(\tilde{t},x,y) = \varphi_0(x,y) - \sqrt{2}\,\sqrt{1 -  
\lambda_1^2 -
\lambda_2^2} \, \log \left(1-\frac{\tilde{t}}{\tilde{t}_0}\right) \,, 
\ees
and for the ``momentum'' constraint, Eq.~(\ref{t1}), we get (redefining  
$\lambda \equiv \lambda_1$ for simplicity)
\bees
\pa_x \lambda - (1-2\,\lambda )\,\pa_x \tilde{\beta}_2 &=& (\pa_x  
\varphi_0)\,
\sqrt{\lambda-\lambda^2} \,,\\
-\pa_y \lambda + (1-2\,\lambda )\,\pa_y \tilde{\beta}_1 &=& (\pa_y  
\varphi_0)\,
\sqrt{\lambda-\lambda^2} \,.
\ees
Note that  the two equations are decoupled and that they
can be solved for $\beta_1$ and $\beta_2$ by quadratures once
$\lambda$ and $\varphi_0$ are given.

\subsection*{Momentum constraints for $d=2$: string frame}
We now re-express the results of the previous section in the string frame
in order to be able to discuss T-duality.
The solutions~(\ref{sol1}) and (\ref{sol2}) for the  two-metric and
the shifted dilaton are
\bees
&& g_{ij}(t,x,y)=e^{2\beta_i(x,y)}\,\delta_{i j}\,
\left(1-\frac{t}{t_0}\right)^{2\,\alpha_i(x,y)}\,,\quad
\alpha_1^2(x,y)+\alpha_2^2(x,y)=1\,, \\
&& \vophi(t,x,y) = \vophi_0(x,y) - \log \left(1-\frac{t}{t_0}\right) \,. 
\ees
In the ``momentum'' constraints, Eq.~(\ref{r4}), the leading terms as $t \rightarrow t_0$
automatically cancel and we get:
\bees
&&\partial_x(\alpha_1\,e^{-\vophi_0})
-e^{-\vophi_0}\,(\alpha_1\,\partial_x\beta_1+
\alpha_2\,\partial_x\beta_2)=0\,, \label{ae1}\\
&&\partial_y(\alpha_2\,e^{-\vophi_0})
-e^{-\vophi_0}\,(\alpha_1\,\partial_y\beta_1+
\alpha_2\,\partial_y\beta_2)=0\,. \label{ae2}
\ees
Introducing the function
\be
\varphi_0(x,y) = \vophi_0(x,y) + \beta_1(x,y) + \beta_2(x,y)
\ee
Eqs.~(\ref{ae1}) and (\ref{ae2}) become
\bees
\label{ab1}
&&
\partial_x\beta_2 = \frac{\partial_x\alpha_1 - \alpha_1  
\partial_x\varphi_0 }
{\alpha_2 - \alpha_1} \,, \\
\label{ab2}
&&  \partial_y\beta_1 = \frac{\partial_y\alpha_2 - \alpha_2
\partial_y\varphi_0 }
{\alpha_1 - \alpha_2} \,.
\ees
Hence, also in the string frame, the ``momentum'' constraints are solved by
quadratures provided   $\varphi_0(x,y)$ and, say,
$\alpha_1$ are used as inputs.
Barring pathologies, we can always change the input
$\alpha_1$ by a duality transformation (e.g. $\alpha_1 \rightarrow
-\alpha_1$) while keeping
$\varphi_0(x,y)$ unchanged, and solve again in $\beta_1$, $\beta_2$,
thus reconstructing the new
$\vophi_0(x,y)$. In this way we will arrive at a rather odd
generalization of
T-duality transformations
in the asymptotic limit of the quasi-homogeneous case. It would be more
natural
to keep
$\vophi_0(x,y)$ rather than $\varphi_0(x,y)$ unchanged under duality,
since this is what happens for the time-dependent parts, but then it
is not clear how a solution can be explicitly constructed.

\end{document}